\documentclass[preprint2,numberedappendix]{emulateapj-rtx4}
\usepackage{graphicx,times,bm,url}
\graphicspath{{./fig/}{./png/}}

\newcommand{\EQ}{\begin{equation}}
\newcommand{\EN}{\end{equation}}
\newcommand{\EQA}{\begin{eqnarray}}
\newcommand{\ENA}{\end{eqnarray}}

\newcommand{\Fig}[1]{Figure~\ref{#1}}

\newcommand{\bra}[1]{\langle #1\rangle}

\newcommand{\mean}[1]{\overline #1}

{}
{}
{}

{}
{}
{}
{}
{}
{}
{}
{}
\newcommand{\meanBB}{\overline{\mbox{\boldmath $B$}}{}}{}
{}
{}
{}
{}
{}
{}
{}
{}

{}
{}
{}

\newcommand{\meanB}{\overline{B}}

{}

{}
{}

\newcommand{\nullvector}{{\bf0}}

\newcommand{\xx}{\bm{x}}

\newcommand{\uu}{\mbox{\boldmath $u$} {}}

\newcommand{\bb}{\mbox{\boldmath $b$} {}}
\newcommand{\BB}{\mbox{\boldmath $B$} {}}

\newcommand{\JJ}{\mbox{\boldmath $J$} {}}

\newcommand{\AAA}{\mbox{\boldmath $A$} {}}

\newcommand{\ff}{\mbox{\boldmath $f$} {}}

\newcommand{\grav}{\mbox{\boldmath $g$} {}}
\newcommand{\nab}{\mbox{\boldmath $\nabla$} {}}

\newcommand{\SSSS}{\mbox{\boldmath ${\sf S}$} {}}

\newcommand{\erf}{{\rm erf}}

\def\Peff{{\cal P}_{\rm eff}}

\def\Pm{\mbox{\rm Pr}_M}
\def\Rm{\mbox{\rm Re}_M}

\def\Rey{\mbox{\rm Re}}

\def\cs{c_{\rm s}}
\def\kf{k_{\it f}} 

\def\urms{u_{\rm rms}}

\def\etatz{\eta_{\rm t0}}

\def\tautd{\tau_{\rm td}}

\def\Beq{B_{\rm eq}}
\def\Beqz{B_{\rm eq0}}

\def\half{{\textstyle{1\over2}}}

\def\onethird{{\textstyle{1\over3}}}

\newcommand{\G}{\,{\rm G}}

\begin{document}

\title{BIPOLAR MAGNETIC STRUCTURES DRIVEN BY STRATIFIED TURBULENCE
  WITH A CORONAL ENVELOPE}

\author{J\"orn Warnecke$^{1,2}$, Illa R. Losada$^{1,2}$, Axel
Brandenburg$^{1,2}$,
Nathan Kleeorin$^{1,3,4}$ and Igor Rogachevskii$^{1,3,4}$
}
\affil{
$^1$Nordita, Royal Institute of Technology and Stockholm University,
Roslagstullsbacken 23, 10691 Stockholm, Sweden\\
$^2$Department of Astronomy, AlbaNova University Center,
Stockholm University, 10691 Stockholm, Sweden\\
$^3$Department of Mechanical
Engineering, Ben-Gurion University of the Negev, POB 653,
Beer-Sheva 84105, Israel\\
$^4$Department of Radio Physics, N.~I.~Lobachevsky State University of
Nizhny Novgorod, Russia
}

\begin{abstract}
We report the spontaneous formation of bipolar magnetic structures
in direct numerical simulations
of stratified forced turbulence with an outer coronal envelope.
The turbulence is forced with transverse random waves only
in the lower (turbulent) part of the domain.
Our initial magnetic field is either uniform in the entire
domain or confined to the turbulent layer.
After about 1--2 turbulent diffusion times,
a bipolar magnetic region of vertical field develops with two coherent
circular structures
that live during one turbulent diffusion time, and then decay
during 0.5 turbulent diffusion times.
The resulting magnetic field strengths inside the bipolar region
are comparable to the equipartition value
with respect to the turbulent kinetic energy.
The bipolar magnetic region forms a loop-like structure
in the upper coronal layer.
We associate the magnetic structure formation with the negative
effective magnetic pressure instability in the two-layer model.
\end{abstract}

\section{Introduction}

Sunspots are visible surface manifestations of magnetic fields inside the Sun.
The magnetic fields in sunspots are sufficiently strong to suppress the transport
of heat via convective motions and therefore they appear as dark on the solar disk.
We are now able to study them with high-resolution
telescopes \citep{SHKR11} as well as with realistic
numerical simulations \citep{Rempel09} to discover new properties of
sunspots.
However, their formation mechanism is still a subject of active
discussion.

It is broadly believed that sunspots are caused by emerging flux
tubes \citep{P55b} that are generated at the bottom
of the convection zone \citep{Par75}.
The magnetic flux tubes become
unstable and rise until they pierce the solar surface to
form bipolar regions and pairs of sunspots \citep{CMS95}.
The rise of flux tubes and the decay of the resulting active regions is also an
ingredient in some dynamo theories to explain the 22 yr
cyclic behavior of the solar magnetic field \citep[e.g.,][]{NC01,Cho03}.
However, these theories assume the existence of thin, isolated magnetic flux
tubes at the bottom of the convection zone
with fields of the order of $10^5\G$.
In direct numerical simulations (DNS)
of solar dynamos, thin flux tubes of comparable
strength have not yet been found \citep[e.g.,][]{GK11,NBBMT13,KMCWB13}.
There is no conclusive evidence of rising magnetic flux tubes
from helioseismology either; see \cite{BBF10,BBLK13},
who also comment on a detection reported by \cite{Ilonidis:2011}.
These difficulties have led to an alternative approach in which sunspots
are formed as shallow phenomena near the solar surface \citep{B05}.

The formation of bipolar regions has recently been found in realistic
radiation-magnetohydrodynamics (MHD) simulations of near-surface convection
\citep{SN12}, were kG magnetic fields were inserted
at the bottom of a $20$\,Mm deep box.
The authors argue that deep downdrafts associated with the supergranulation
concentrate the magnetic field and determine thereby the scale of their
separation.
They also emphasize the importance of radiative cooling at
the surface.

Two alternative mechanisms for
the spontaneous formation of flux concentrations have been
discussed in the literature.
One is based on a turbulent thermo-magnetic instability
in small-scale turbulence with radiative boundaries,
where the magnetic quenching of the turbulent convective heat
transport is used
in mean-field simulations (MFSs) to produce
flux concentrations \citep{KM00}.
The other mechanism is based on the suppression of the total
turbulent pressure
(the sum of hydrodynamic and magnetic contributions)
by the magnetic field.
This leads to the so-called negative effective magnetic pressure
instability (NEMPI), which is another mechanism that forms flux
concentrations in a stratified turbulent medium \citep[see,
e.g.,][]{BKKMR11,KBKMR12}.
The original idea of NEMPI goes back to early work by
\cite{Kleeorin89,KRR1990} and has led to numerous studies both
analytically \citep{KR1994,KMR96,RK2007} and through DNS and MFS
\citep{BKR10,BKKR12,LBKMR2012,LBKR2013a,LBKR2013b,JBKMR13}.

More recently, \cite{BKR2013} found
super-equipartition magnetic field concentrations in DNS
by imposing a relatively weak vertical magnetic field
on forced strongly stratified turbulence.
Their results may be related to
earlier simulations of turbulent convection in which a vertical
magnetic field was found to segregate into magnetized and unmagnetized regions
\citep{Tao_etal98}, but it is still unclear whether
those results are also caused
by NEMPI or by some other mechanism that still remains to be understood.
However, neither the turbulent thermo-magnetic instability
nor NEMPI have yet been able to produce bipolar magnetic regions.
Except for the work of \cite{SN12}, bipolar regions
have previously been studied by advecting a semi-torus shaped twisted flux tube of 9\,kG
through the bottom boundary at a 7.5\,Mm depth of thermally relaxed
convection \citep{Cheung_etal10}.

In the current study
we demonstrate that bipolar regions can be obtained naturally
in DNS in the presence of an outer coronal layer.
Indeed, all previous simulations of NEMPI and the turbulent
thermo-magnetic instability have a rigid upper boundary,
which is clearly unrealistic.
In this Letter we alleviate the constraints from a rigid upper boundary
by combining NEMPI
in a forced turbulent layer
with a coronal envelope at the top.
This approach has been
successful in generating plasmoids reminiscent of coronal mass
ejections \citep{WBM11}.
In that work, a simplified corona is combined
with a large-scale dynamo that is
either generated by forced turbulence
in a Cartesian domain \citep{WB10} or
in spherical coordinates \citep{WBM11}, or through
self-consistent turbulent convection \citep{WKMB12}.
This approach leads to a more realistic ``boundary condition'' at the interface
between the dynamo and corona without using, for example, radiative transfer.
Not only has it led to coronal ejections, but also to the discovery of a
change of sign of magnetic helicity some distance away from the star
\citep{WBM12}, and the realization
that both the dynamo and the differential rotation may be affected by the presence
of a free boundary as modeled by the coronal envelope \citep{WKMB13b}.

For a more realistic model, several other factors need to be accounted for.
Except for radiation and ionization that would potentially lead to
new effects \citep[see, e.g.,][]{BB13}, which might distort our
interpretation, there would be the
issue of the origin of the magnetic field itself.
It should really come from a self-consistent
large-scale dynamo beneath the surface.
Again, this leads to new complications, through its interaction
with NEMPI, which have only recently been addressed
by \cite{JBKMR13} and \cite{LBKR2013a}.
To avoid of those complications, we take a horizontal magnetic field
as the initial condition.
We mainly consider the case of a uniform magnetic field
that is then also present in the corona at all times (case~A),
and compare with a case in which it is initially only in
the turbulence layer, but slowly decaying (case~B).
These fields are weak, but they might have interesting unexpected and
unexplored effects that will be the topic of the current Letter.
At this point we cannot be certain that it has direct applications to the Sun,
but in view of the fact that research into NEMPI has been growing rapidly,
it is important to explore all aspects of it before we can make definitive
statements about its applicability to sunspots.

\section{The model}

Our model is similar to that of \cite{BKR2013},
where they solve the isothermal hydromagnetic equations in the presence of
vertical gravity $\grav=(0,0,-g)$, but here the random nonhelical forcing $\ff$
acts just in the lower part ($-\pi<z<0$) of a Cartesian domain of size
$L_x\times L_y\times L_z$, where $L_x=L_y=2\pi$ and $L_z=3\pi$.
Also, instead of a vertical field, we now take a weak uniform
horizontal magnetic field, either as an imposed one throughout the domain,
$\BB_{\rm imp}=(0,B_0,0)$ with $\AAA=\nullvector$ (case~A),
or as an initial one in the lower part ($z<0$) by putting
$A_x=-\max(0,-z)\,B_0$ and $\BB_{\rm imp}=\nullvector$ (case~B).
Similar to \cite{WB10}, we employ a forcing function $\ff(\xx,t)$
that is modulated by
\begin{equation}
\theta_w(z)=\half\left(1-\erf{z\over w}\right),
\end{equation}
where $w$ is the width of the transition.
We thus solve the equations for the velocity $\uu$,
the magnetic vector potential $\AAA$, and the density $\rho$:
\begin{equation}
{D\uu\over D t}=\grav+\theta_w(z)\ff
+{1\over\rho}[-\cs^2\nab\rho+\JJ\times\BB+\nab\cdot(2\nu\rho\SSSS)],
\end{equation}
\begin{equation}
{\partial\AAA\over\partial t}=\uu\times\BB+\eta\nabla^2\AAA,
\end{equation}
\begin{equation}
{D \ln\rho\over D t}=-\nab\cdot\uu,
\end{equation}
where $\nu$ is the kinematic viscosity, $\eta$ is the magnetic diffusivity,
$\BB=\BB_{\rm imp}+\nab\times\AAA$ is the magnetic field,
$\JJ=\nab\times\BB/\mu_0$ is the current density,
$\mu_0$ is the vacuum permeability,
${\sf S}_{ij}=\half(u_{i,j}+u_{j,i})-\onethird\delta_{ij}\nab\cdot\uu$
is the traceless rate of strain tensor, and commas denote
partial differentiation.
The forcing function $\ff$ consists of
random, white-in-time, plane non-polarized waves \citep{Hau04} with an
average wavenumber $\kf=30\,k_1$,
where $k_1=2\pi/L_x$ is the lowest wavenumber in the domain
in the horizontal direction.
The forcing strength is such that the turbulent rms velocity is approximately
independent of $z$ with $\urms=\bra{\uu^2}^{1/2}\approx0.1\,\cs$.
The value of the gravitational acceleration $g$ is
related to the density scale height $H_\rho=\cs^2/g$ and is chosen such that
$k_1 H_\rho=1$, which leads to a density contrast between bottom and top
of $\exp(3\pi)\approx1.2\times10^4$.

Except for the profile of the forcing function
and the extent of the domain, our setup is also similar to
\cite{KBKMR12}, who used the same values of
the fluid Reynolds number $\Rey\equiv\urms/\nu\kf=38$,
the magnetic Prandtl number $\Pm\equiv\nu/\eta=1/2$, and thus of
the magnetic Reynolds number $\Rm\equiv\Rey\,\Pm=19$,
which is known to lead to negative effective magnetic pressure
for the mean magnetic fields, $\meanBB$,
in the range $0<|\meanBB|/\Beq<0.4$.
Here, $\meanBB$ is obtained by averaging over the scale
of several turbulent eddies.
The magnetic field is expressed in units of the local
equipartition field strength, $\Beq=\sqrt{\mu_0\rho} \, \urms$,
while the initial magnetic field $B_0$
is specified in units of the value at $z=0$,
namely $\Beqz=\sqrt{\mu_0\rho_0} \, \urms$, where $\rho_0=\rho(z=0)$.
We use $B_0/\Beqz=0.02$, which is
also the field strength used in the main run of \cite{BKR2013}.
Time is expressed in turbulent-diffusive times,
$\tautd=(\etatz k_1^2)^{-1}$, where $\etatz=\urms/3\kf$ is
the estimated turbulent magnetic diffusivity.

\begin{figure}[t!]
\begin{center}
\includegraphics[width=0.49\columnwidth]{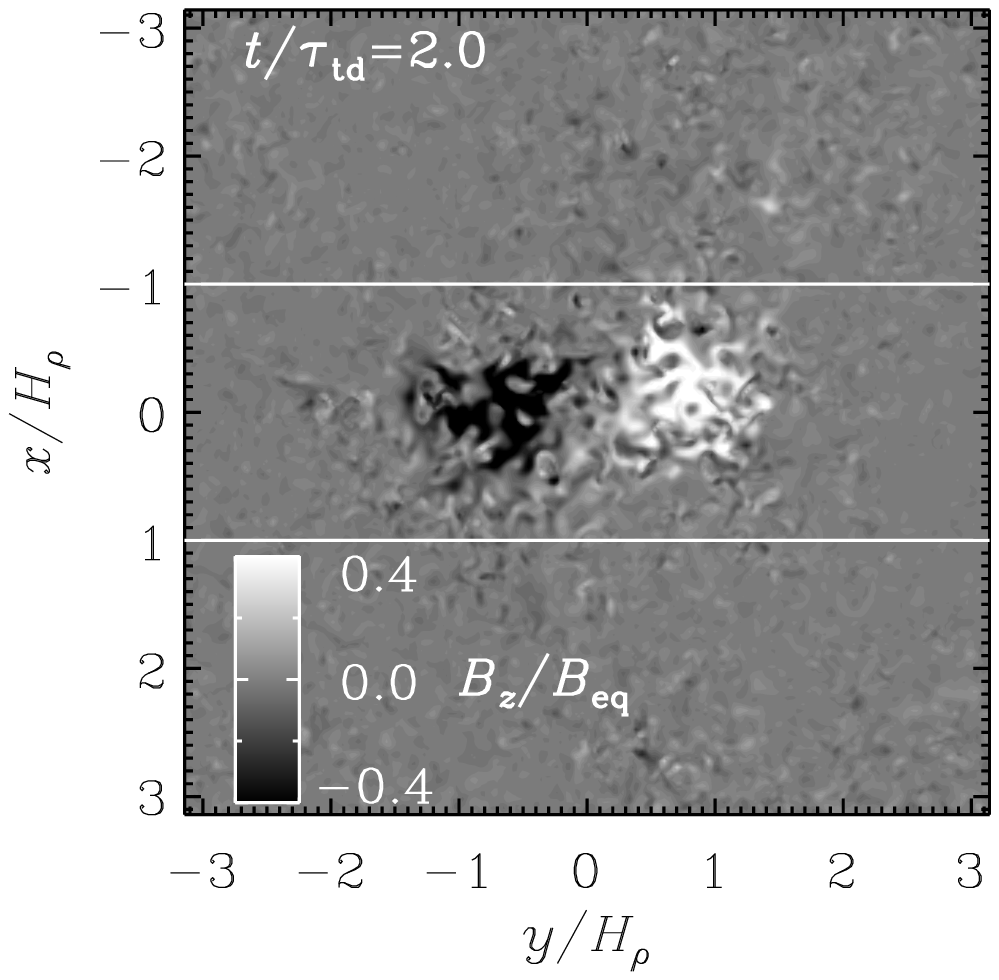}
\includegraphics[width=0.49\columnwidth]{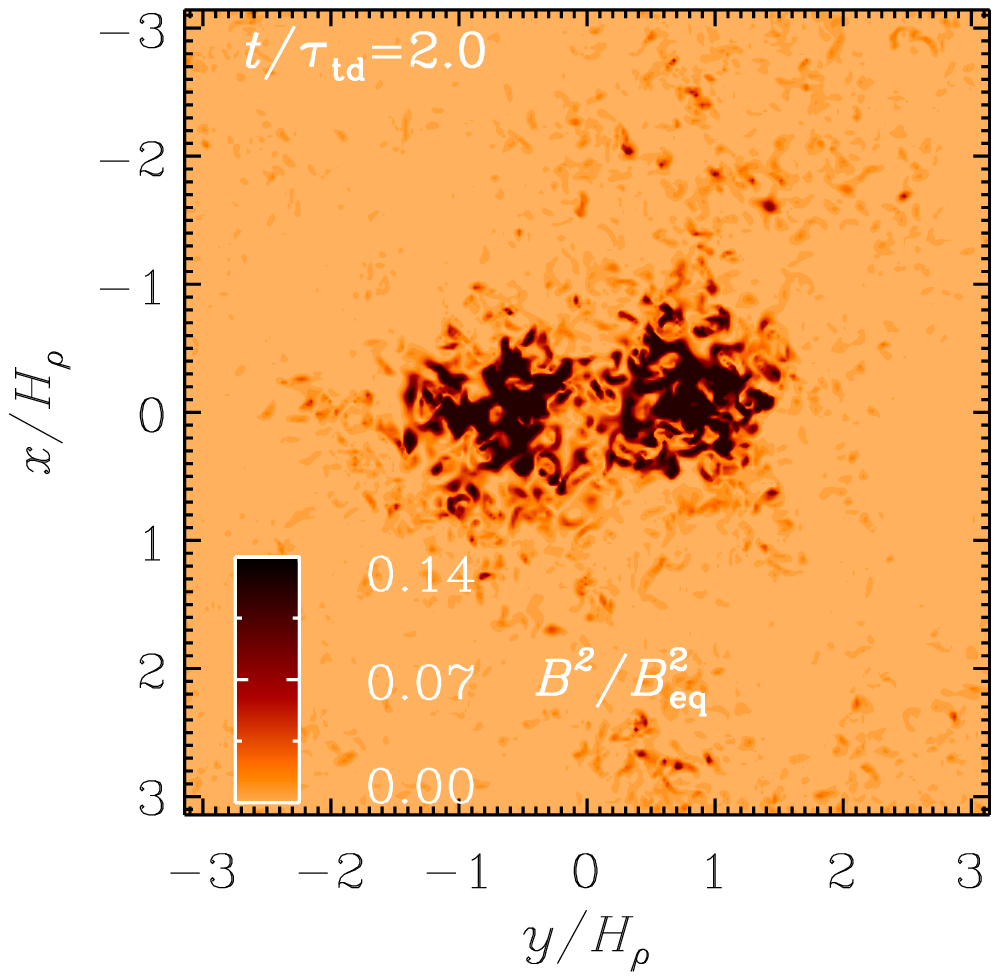}
\end{center}\caption[]{
Left panel: normalized vertical magnetic field $B_z/\Beq$ of the bipolar region at
the surface ($z=0$) of the simulation domain.
The white lines delineate the area shown in \Fig{ppslices_xy}.
Right panel: normalized magnetic energy $\BB^2/\Beq^2$ of the
two regions relative to the rest of the surface.
Note that we clip both color tables to increase the contrast of
the structure. The field strength reaches around $B_z/\Beq=1.4$.
Case~A.
}\label{b2_xy}
\end{figure}

\begin{figure}[t!]
\begin{center}
\includegraphics[width=\columnwidth]{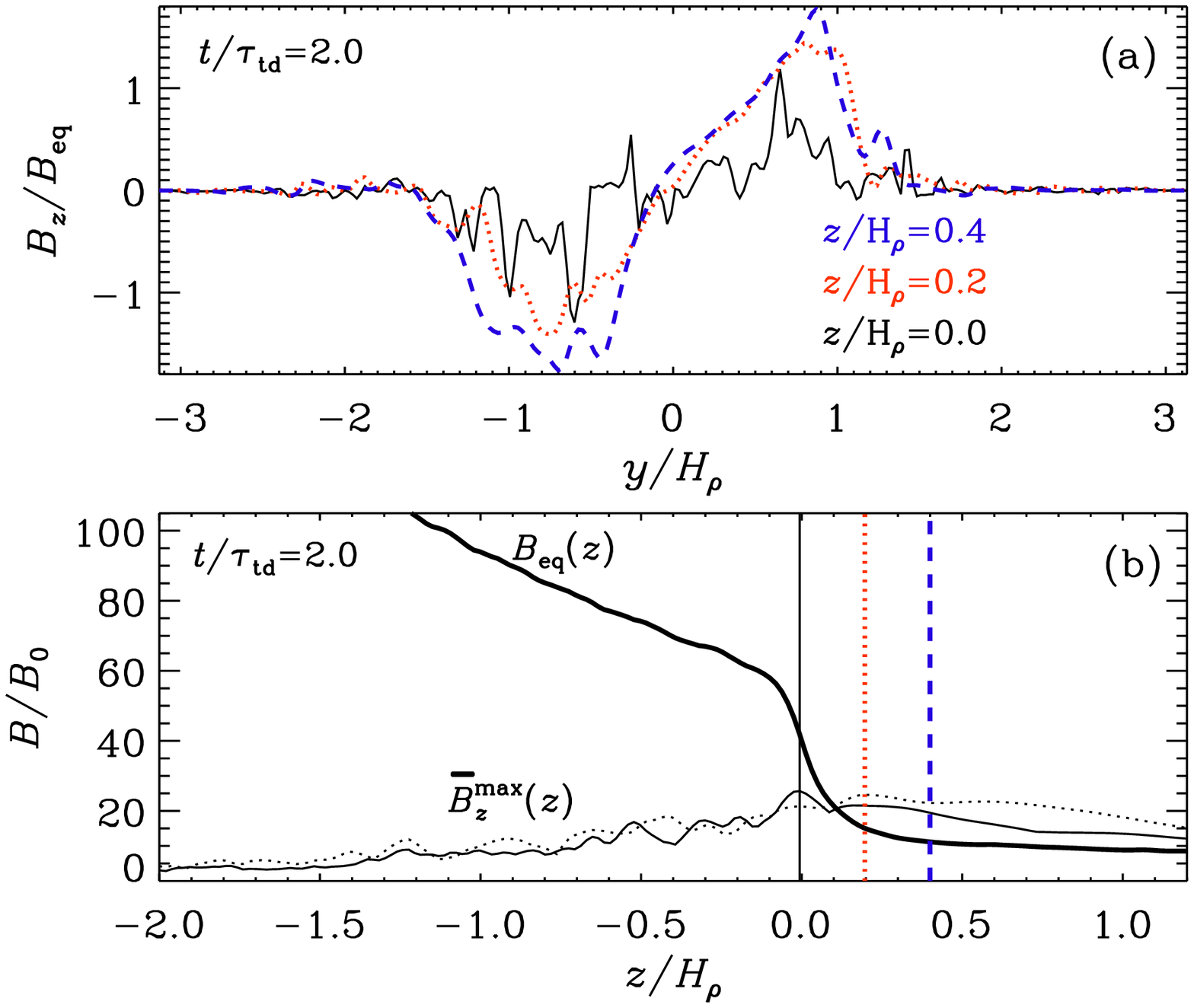}
\includegraphics[width=\columnwidth]{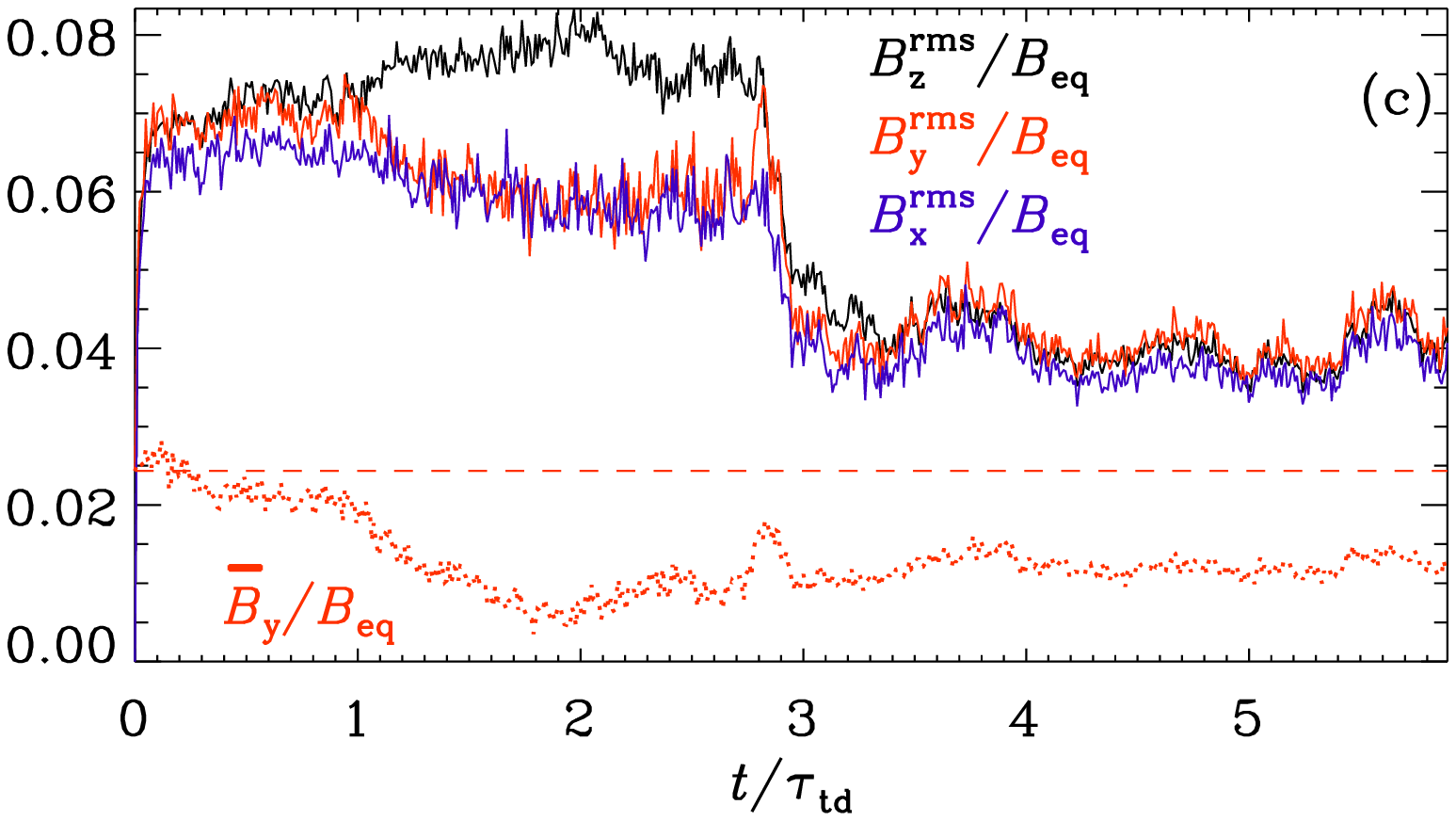}
\includegraphics[width=\columnwidth]{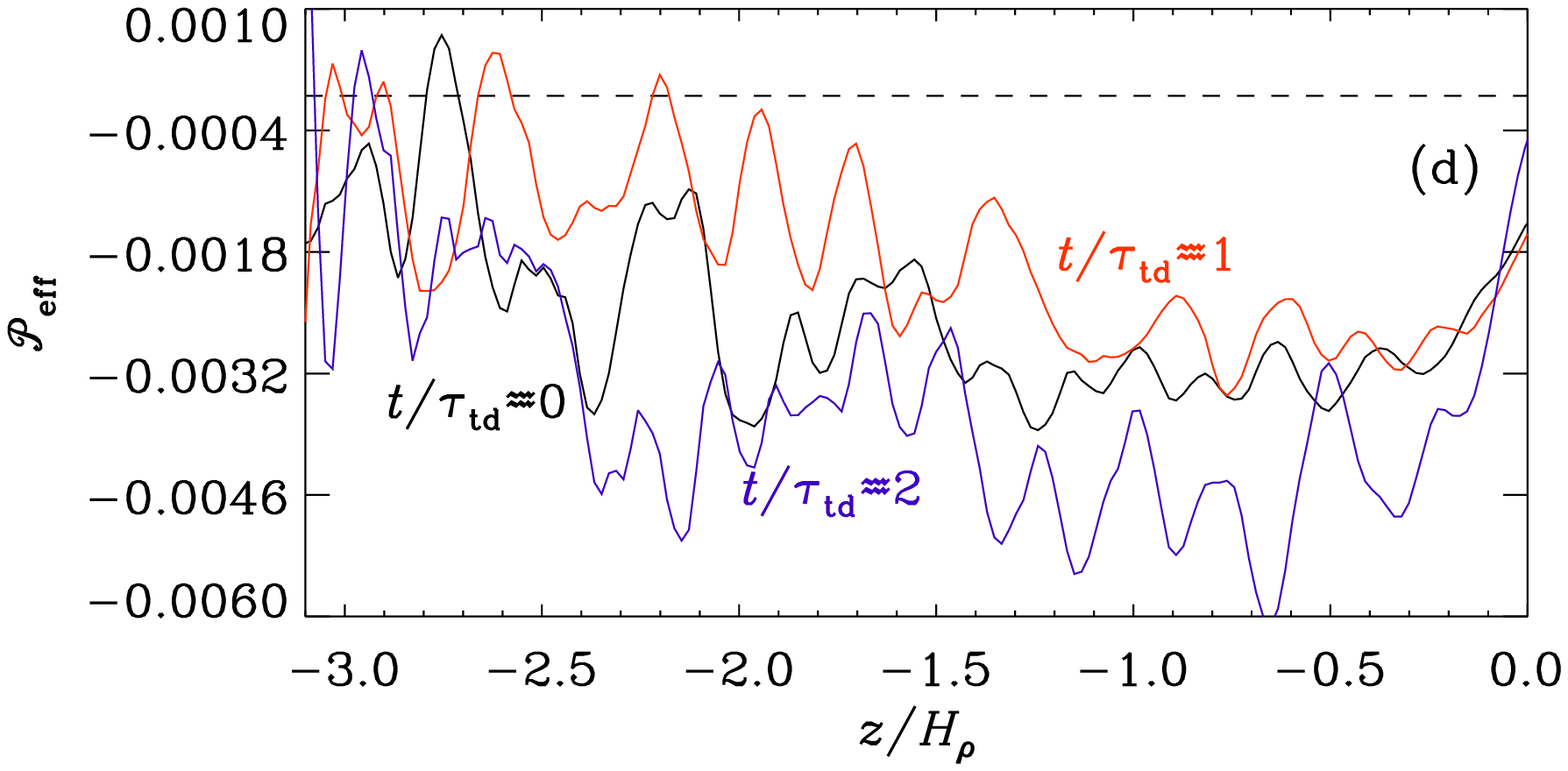}
\end{center}\caption[]{
(a)
Profiles of $B_z(y)/\Beq$ through $x=0$ at three heights ($z/H_\rho=0, 0.2, 0.4$),
whose values are indicated also by vertical bars in the next panel.
(b)
Vertical profiles of $\Beq(z)/B_0$ (thick solid line) and
$\meanB_z^{\max}(z)/B_0$,
shown as a regular solid line at $t/\tautd=2.0$.
The dotted lines show the profiles of $\meanB_z^{\max}(z)/B_0$,
which are similar to those of $-\meanB_z^{\min}(z)/B_0$.
(c)
Growth and decay of the different magnetic field
components at the surface ($z$=0).
The solid lines represent the rms of the vertical field
$B_z^{\rm rms}=\bra{B_z^2}_{xy}$ (black),
and the horizontal $B_y^{\rm rms}=\bra{B_y^2}_{xy}$ (red),
$B_x^{\rm  rms}=\bra{B_x^2}_{xy}$ (blue).
The horizontal averaged magnetic field in the direction of the initial
field $\mean{B_y}=\bra{B_y}_{xy}$ is shown with a dotted red line,
the value of the initial field is indicated by a dashed red line.
All values are normalized by the equipartition field strength at the
surface.
(d)
Effective magnetic pressure $\Peff$ over height $z$.
The three lines indicate different times during
the simulation: $t/\tautd\approx0$ (black), $t/\tautd\approx1$ (red)
and $t/\tautd\approx2$ (blue).
The pressure is averaged over a time interval
$\Delta t/\tautd\approx0.15$ and smoothed over four grid points.
The dashed line represents the zero line.
Case~A.
}\label{pBBeq_one}
\end{figure}

\begin{figure*}[t!]
\begin{center}
\includegraphics[width=.97\textwidth]{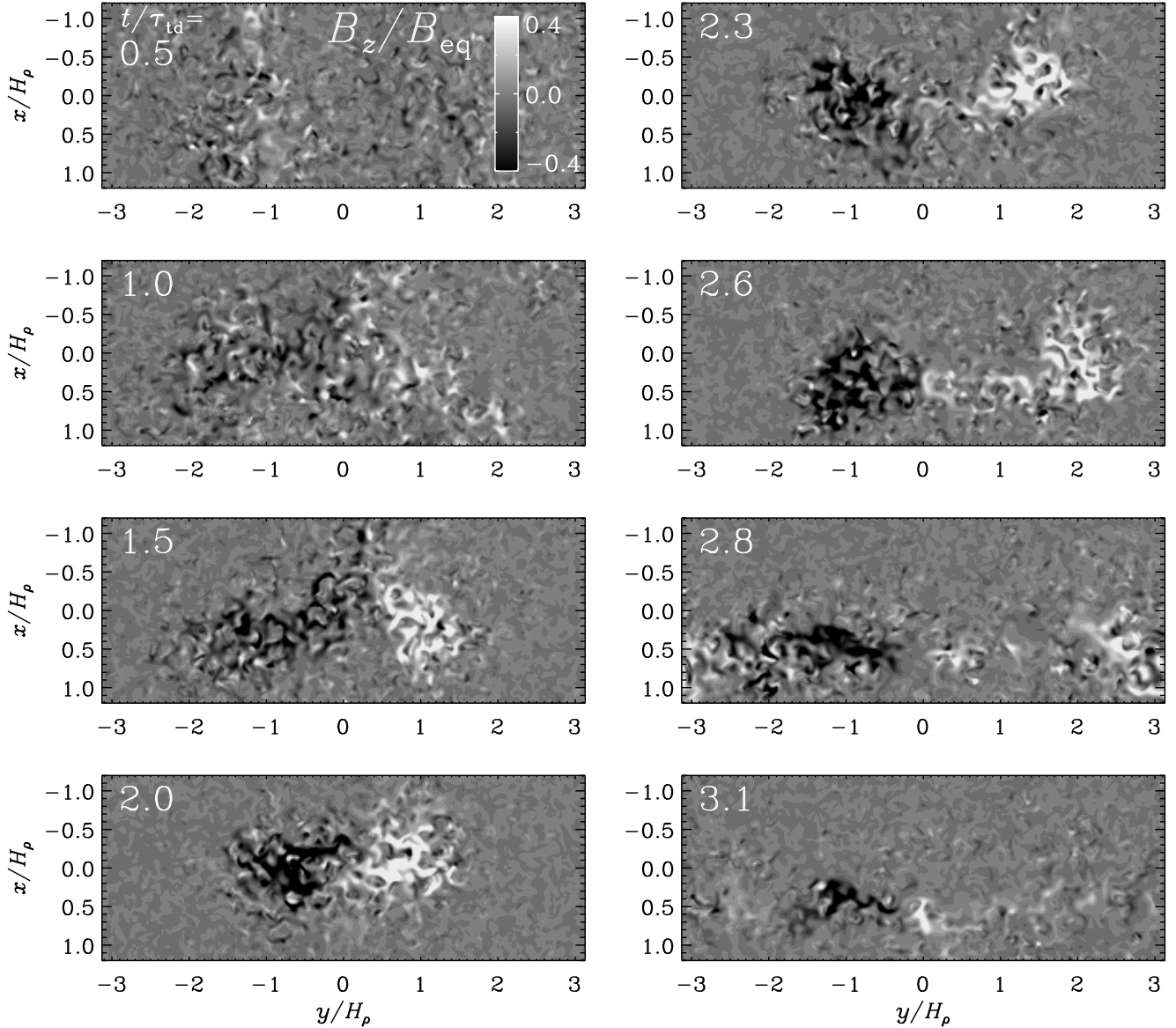}
\end{center}\caption[]{

Visualizations of horizontal cross-sections of $B_z(x,y)/\Beq$
through $z=0$, for case~A, similar to \Fig{b2_xy},
but only data in $-1.2\leq x/H_\rho\leq1.2$ are shown.
}\label{ppslices_xy}
\end{figure*}

\begin{figure*}[t!]
\begin{center}
\includegraphics[width=0.49\textwidth]{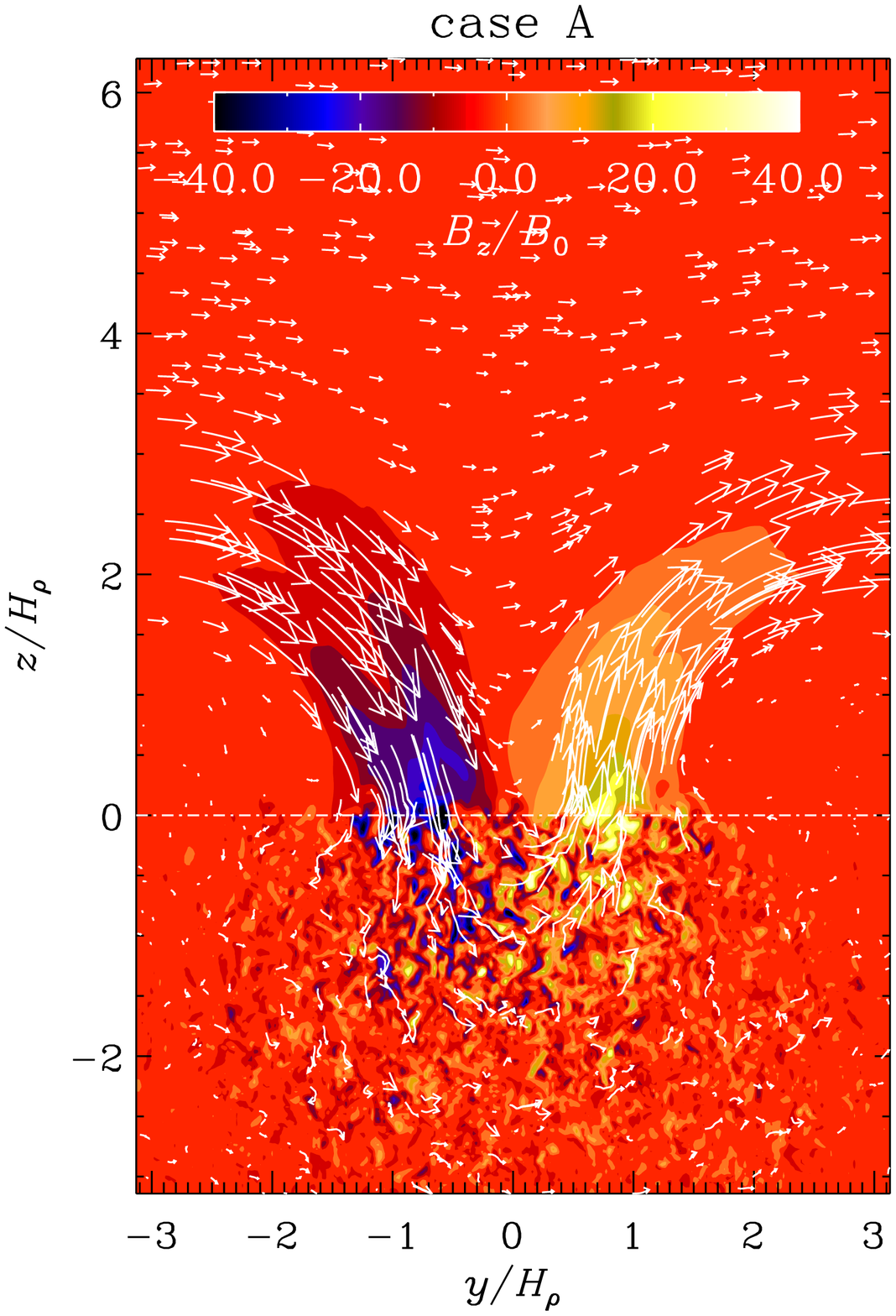}
\includegraphics[width=0.49\textwidth]{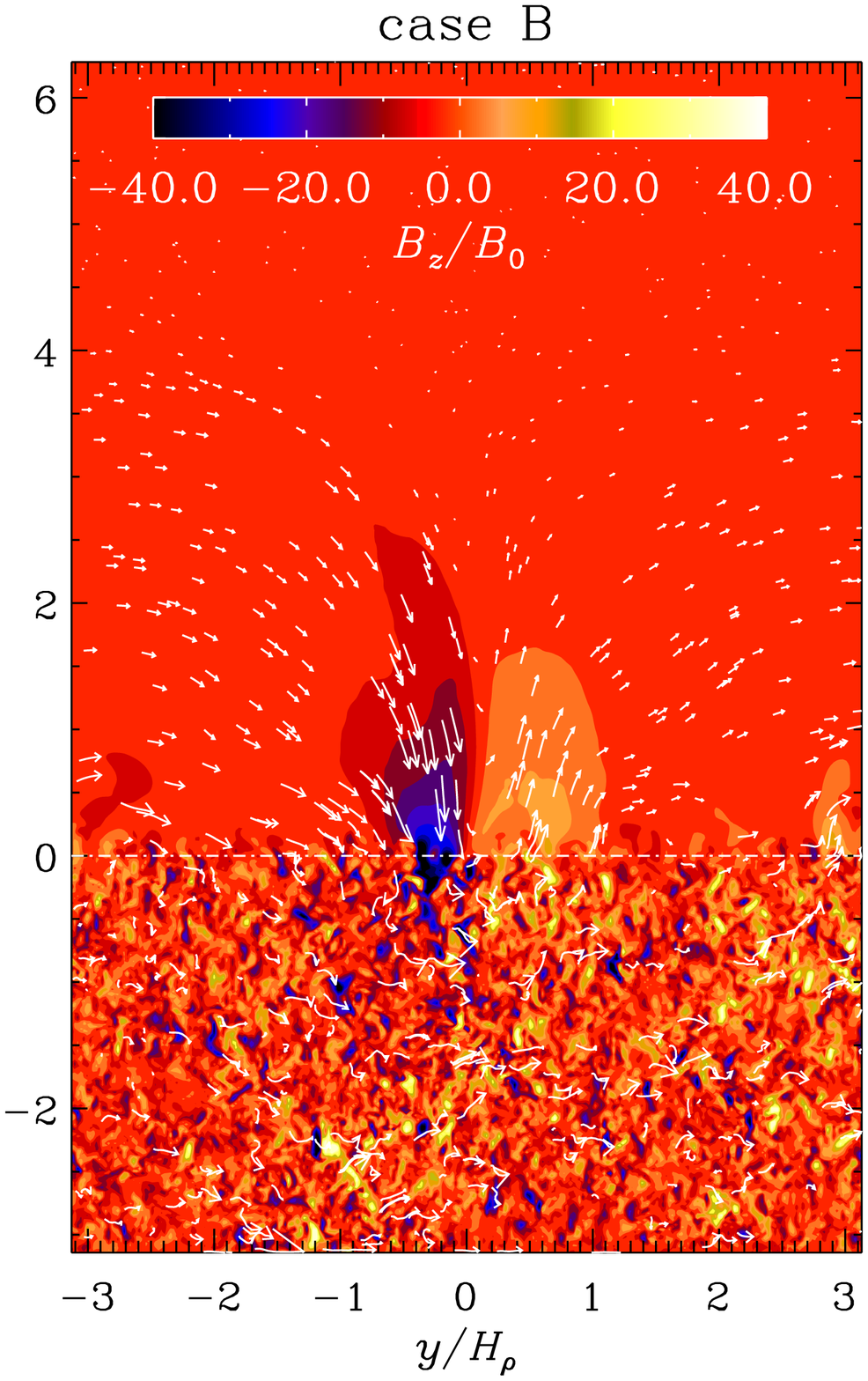}
\end{center}\caption[]{

Visualizations of vertical cross-sections of $B_z(y,z)/B_0$
together with magnetic field vectors in the $yz$ plane
through the $x$ location of the flux convergence for case~A at
$t/\tautd=2.0$ (left) and case~B at $t/\tautd=1.83$ (right).
The dash-dotted lines indicate the surface at $z=0$.
}\label{ppslice_yz}
\end{figure*}

\begin{figure*}[t!]
\begin{center}
\includegraphics[width=\textwidth]{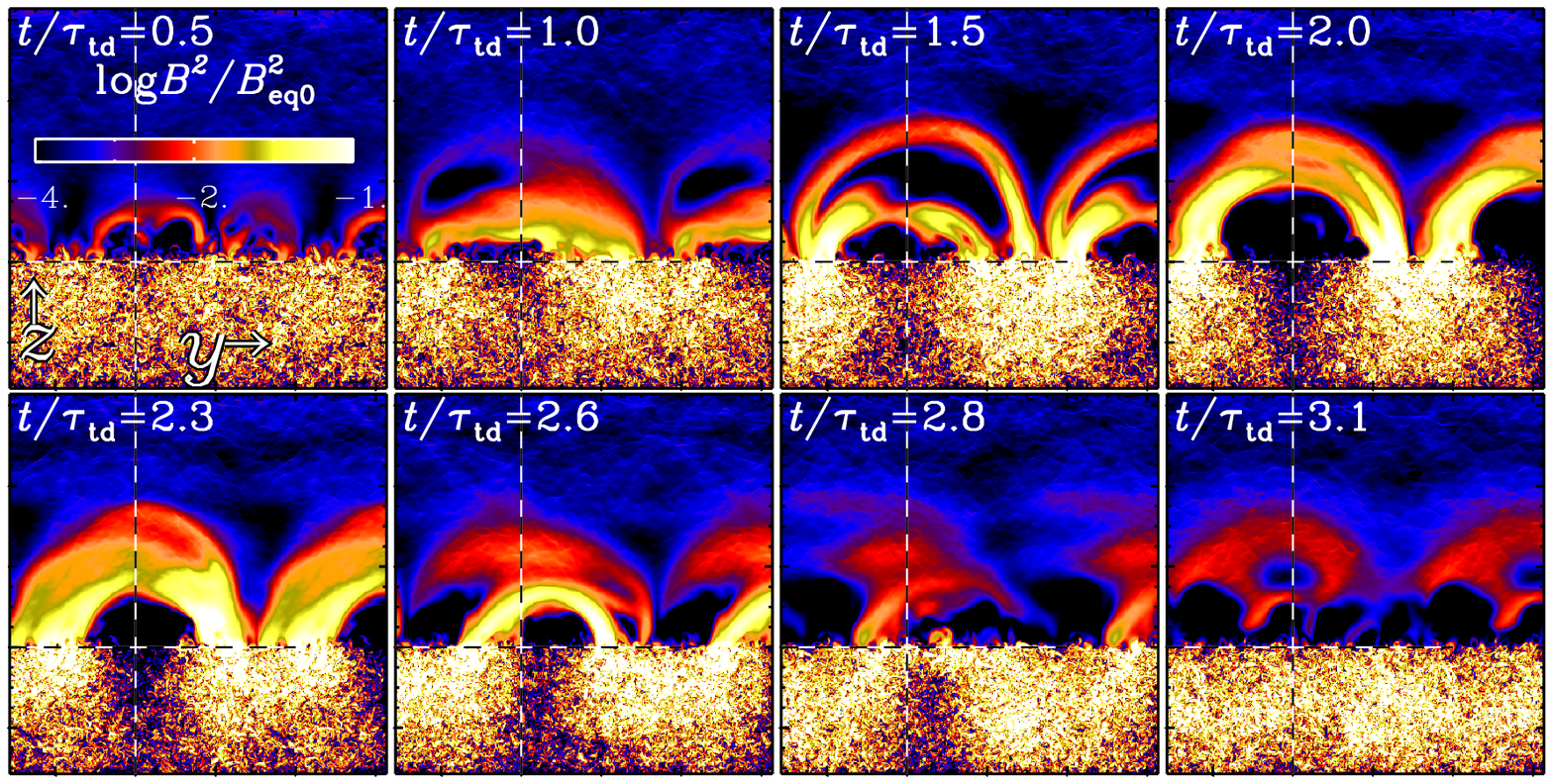}
\end{center}\caption[]{
Time series of normalized magnetic energy density $\BB^2/\Beqz^2$ in a vertical
cut through the bipolar region at $x=0$ for case~A.
The domain has been replicated by 50\% in the $y$ direction to give
a more complete impression about spot separation and arch length.
The black-white dashed lines mark the surface ($z=0$) and the replicated part.
}\label{b2_yz}
\end{figure*}

The simulations are performed with the {\sc Pencil Code},%
\footnote{{\tt http://pencil-code.googlecode.com}}
which uses sixth-order explicit finite differences in space and a
third-order accurate time stepping method.
We use a resolution of $256\times256\times512$ mesh points
in the $x$, $y$, and $z$ directions.
We adopt periodic boundary conditions in the $xy$ plane and present our results
by shifting our coordinate system
such that the regions of interest lie at the center around $x=y=0$.
On $z=-\pi$ we apply a stress-free perfect conductor condition
and on $z=2\pi$ a stress-free vertical field condition.

\section{Results}

We report the spontaneous formation and decay of bipolar magnetic regions
at the surface ($z=0$), which is
the boundary between regions with and without forcing.
These two parts resemble the upper convection zone and a simplified corona of
the Sun.
In \Fig{b2_xy}, we show for case~A the bipolar region as the normalized
vertical magnetic field $B_z/\Beq$ (left panel) and the normalized
magnetic energy $\BB^2/\Beq^2$ (right panel) at the moment of maximum
strength,
$t/\tautd=2$.
Note that the $y$ direction points to the right,
while the positive $x$ direction points downward, so the coordinate
system has been rotated by $90^\circ$ to allow a view that is more similar
to how bipolar regions are oriented on the solar disk.
The shapes of both magnetic spots are nearly circular, but are still
disturbed by the turbulent motion acting on the magnetic field.
The field outside this bipolar region is weak: almost all the magnetic
field is concentrated inside the bipolar region.
At and slightly above $z=0$, we find field strengths significantly above
the equipartition value.
This is seen more clearly in \Fig{pBBeq_one}(a), where we show
for case~A profiles of $B_z(y)/\Beq$ through $x=0$ at three heights.
We normalize the magnetic field
by its local equipartition value, $\Beq(z)$ (\Fig{pBBeq_one}a)
or by the strength of the initial field $B_0$ (\Fig{pBBeq_one}b).
The field $\Beq(z)$ is also shown as a thick line in \Fig{pBBeq_one}(b).
At the surface we have $\Beq(0)/B_0\approx50$, but for $z>0$
it drops sharply to values below 20.
At each height, we have computed maximum and minimum field strengths
of the mean field, $\meanBB$, defined here by Fourier filtering
to include only wavenumbers below $\kf/2$,
as functions of $z$, i.e., $\meanB_z^{\max}(z)$ and
$-\meanB_z^{\min}(z)$, respectively.
It turns out that $\meanB_z^{\min}(z)\approx-\meanB_z^{\max}(z)$, and that
both decline more slowly with height than $\Beq$, so the field reaches
super-equipartition strength in the outer parts.

We recall that in case~A,
we have $\BB_0=\BB_{\rm imp}$ over the whole domain.
However, the field quickly becomes tangled by the random velocity field
in $z<0$ where the forcing is acting to
produce small-scale magnetic fields on the scale of the turbulence.
In the upper part, however, this horizontal field stays roughly unchanged
up to the instant
when it becomes affected by a large-scale instability (NEMPI).
As shown in \Fig{pBBeq_one}(c), the rms values of
the three components of the magnetic field at the surface ($z=0$)
grow rapidly until they saturate at $t/\tautd \approx 0.2$.
At $t/\tautd\approx1$, the magnetic field has attained a strong
vertical component while the horizontal one declines.
By the time $t/\tautd=2$, the vertical field is stronger
than the horizontal field until all three components decay rapidly to
a lower value and saturate there.

To see whether the increase of vertical magnetic field and
structure formation is related to NEMPI, we show in
\Fig{pBBeq_one}(d) that the effective magnetic pressure
$\Peff$ is indeed negative below the surface in the turbulent region
for case~A.
It can be calculated
using an approach applied by \cite{BKKR12}:
\begin{equation}
\Peff={\meanBB^2\over2\Beq^2}+{\left(\Pi_{xx}^{(B)}-\Pi_{xx}^{(0)}
+\Pi_{zz}^{(B)}-\Pi_{zz}^{(0)}\right)\over\Beq^2},
\end{equation}
where $\Pi_{ij}^{(B)}\equiv\overline{\rho u_iu_j}
+\half\delta_{ij}\mu_0^{-1}\overline{\bb^2}-\mu_0^{-1}\overline{b_ib_j}$
and $\Pi_{ij}^{(0)}$ are the components
of the total (Reynolds and Maxwell) turbulent stress tensor with
and without mean magnetic influence, respectively,
and $\delta_{ij}$ is the unit Kronecker tensor.

Horizontal cross-sections of $B_z(x,y)/\Beq$ through $z=0$
are shown in \Fig{ppslices_xy} for case~A at different times.
At $t/\tautd=1$, structures begin to form that become more coherent
and more circular while
decreasing their distance to a minimum until they lie directly next to each
other ($t/\tautd=2$).
This is also the time of maximum field strength and maximum coherence, and
agrees with the peak of $B_z^{\rm rms}$ in \Fig{pBBeq_one}(c).
After this time, the distance and field strength of the
two polarities decrease until no large-scale structures are visible ($t/\tautd=3.5$).

In \Fig{ppslice_yz} we compare $yz$ slices through the bipolar region
for cases~A and B.
The vertical magnetic field is color coded and arrows indicate
magnetic field vectors in the plane.
The field in the region of $z>3$ is not strongly disturbed by the structure
formation and represents $B_{\rm imp}$ in case~A and is $\approx0$ in
case~B.
The weaker field concentration in case~B might be related to the
decay of the initial field.
In both cases, the bipolar spot orientation is peculiar
because an emerging flux tube with similar field
direction as the imposed field would cause an inverted vertical flux
configuration.

The formation of magnetic structures
can be caused by the negative contribution of turbulence to the
effective large-scale magnetic pressure (the sum of turbulent and
non-turbulent contributions).
For large magnetic Reynolds numbers, the turbulent contributions
are larger than the non-turbulent ones, and the
effective magnetic pressure becomes negative; see \Fig{pBBeq_one}(d).
This results in a large-scale instability, which
causes a redistribution of mass so that a large-scale flow is generated.
This flow may also drive the magnetic field patches together.
Since turbulence has produced similar strengths for
all three components of the magnetic field, and since NEMPI allows for
stronger vertical fields
than horizontal ones \citep{BKR2013}, the result is the
formation of strong vertical field structures.

It is important to realize that our setup corresponds to an
initial value problem in the sense that the magnetic field affects
the effective magnetic pressure.
It changes the horizontally symmetric
background state, which is unstable
with respect to NEMPI.
This leads to the formation of magnetic structures that tend to stabilize
the system.
This is the reason why, with our present setup, a bipolar magnetic region
occurs only once.
Of course, if we apply this mechanism to the Sun, the imposed magnetic
field would be provided by a dynamo acting in the convection zone,
which would certainly show cycles and fluctuations.

In the coronal region, magnetic loops form that connect the far ends
of spots of opposite polarity, as can be seen by plotting
1.5 times the full periodic length in the $y$ direction.
In \Fig{b2_yz} we show for case~A the temporal evolution of the normalized
magnetic energy density $\BB^2 /\Beq^2$ in such a $yz$ slice.
The large-scale loop
is strongest at $t/\tautd\approx2$ when the two magnetic spots are most concentrated.
An exploratory run with a larger extent in the $y$-direction
resulted in similarly oriented bipolar structures with somewhat
larger separation.

\section{Conclusions}

The current study has shown
for the first time that a bipolar magnetic region can emerge
and later decay as a natural consequence of stratified turbulence with
a coronal envelope and a sufficiently large domain size.
For the formation process of the bipolar region, radiative cooling
and a particular large-scale flow topology as in \cite{SN12}
seem to be unimportant.
However, they will certainly play a role in determining the detailed
structure of sunspots \citep{Rempel09,Cheung_etal10}.
The interaction with a weak pre-existing coronal field is not unrealistic.
A similar interaction has previously been employed
in connection with the production of X-ray jets and flares
\citep{Shibata05}.
It supports the formation of a bipolar region (case~A),
but is otherwise not critical (case~B).
On the other hand, preliminary calculations with a potential
field boundary condition and no corona have not yet resulted
in bipolar regions.

We find the two polarities approaching each other in the forming
process and separating in the decaying phase.
This is an important aspect that was not previously expected
from turbulent processes \citep{Spruit12}.
The observed formation of the bipolar regions in cases~A and B is consistent
with NEMPI, which predicts that, even though the optimal field
strength for the excitation of NEMPI is significantly below the
equipartition value, the magnetic field strengths inside the
bipolar region can be significantly above the equipartition value of the
turbulence at and slightly above the surface \citep{BGJKR13}.

Useful extensions of the model include more realistic stratifications
such as a polytropic one in the lower turbulent
part \citep{LBKR2013b} combined with a sharp drop in temperature
in the upper layer caused by ionization and radiation.
Furthermore, above the chromosphere the temperature should increase again
in the solar corona \citep[cf.][]{WKMB13b}.

\acknowledgments

We thank the referee for useful comments and
acknowledge the computing resources provided by the Swedish
National Allocations Committee
at the Center for Parallel Computers at the Royal Institute of Technology in
Stockholm and the High Performance Computing Center North in Ume{\aa}.
This work was supported in part by the European Research Council
under the AstroDyn Research Project No.\ 227952,
by the Swedish Research Council under the project grants
621-2011-5076 and 2012-5797, by EU COST Action MP0806,
by the European Research Council under the Atmospheric Research Project No.\
227915, and by a grant from the Government of the Russian Federation under
contract No. 11.G34.31.0048.

\bibliography{paper}

\begin{thebibliography}{42}
\expandafter\ifx\csname natexlab\endcsname\relax\def\natexlab#1{#1}\fi

\bibitem[{{Barekat} \& {Brandenburg}(2013)}]{BB13}
{Barekat}, A., \& {Brandenburg}, A. 2013, \aap, submitted (arXiv:1308.1660)

\bibitem[{{Birch} {et~al.}(2010){Birch}, {Braun}, \& {Fan}}]{BBF10}
{Birch}, A.~C., {Braun}, D.~C., \& {Fan}, Y. 2010, \apjl, 723, L190

\bibitem[{{Birch} {et~al.}(2013){Birch}, {Braun}, {Leka}, {Barnes}, \&
  {Javornik}}]{BBLK13}
{Birch}, A.~C., {Braun}, D.~C., {Leka}, K.~D., {Barnes}, G., \& {Javornik}, B.
  2013, \apj, 762, 131

\bibitem[{{Brandenburg}(2005)}]{B05}
{Brandenburg}, A. 2005, \apj, 625, 539

\bibitem[{{Brandenburg} {et~al.}(2013{\natexlab{a}}){Brandenburg}, {Gressel},
  {Jabbari}, {Kleeorin}, \& {Rogachevskii}}]{BGJKR13}
{Brandenburg}, A., {Gressel}, O., {Jabbari}, S., {Kleeorin}, N., \&
  {Rogachevskii}, I. 2013{\natexlab{a}}, A\&A, submitted (arXiv:1309.3547)

\bibitem[{{Brandenburg} {et~al.}(2011){Brandenburg}, {Kemel}, {Kleeorin},
  {Mitra}, \& {Rogachevskii}}]{BKKMR11}
{Brandenburg}, A., {Kemel}, K., {Kleeorin}, N., {Mitra}, D., \& {Rogachevskii},
  I. 2011, \apjl, 740, L50

\bibitem[{{Brandenburg} {et~al.}(2012){Brandenburg}, {Kemel}, {Kleeorin}, \&
  {Rogachevskii}}]{BKKR12}
{Brandenburg}, A., {Kemel}, K., {Kleeorin}, N., \& {Rogachevskii}, I. 2012,
  \apj, 749, 179

\bibitem[{{Brandenburg} {et~al.}(2010){Brandenburg}, {Kleeorin}, \&
  {Rogachevskii}}]{BKR10}
{Brandenburg}, A., {Kleeorin}, N., \& {Rogachevskii}, I. 2010, AN, 331, 5

\bibitem[{{Brandenburg} {et~al.}(2013{\natexlab{b}}){Brandenburg}, {Kleeorin},
  \& {Rogachevskii}}]{BKR2013}
{Brandenburg}, A., {Kleeorin}, N., \& {Rogachevskii}, I. 2013{\natexlab{b}},
  \apjl, 776, L23

\bibitem[{{Caligari} {et~al.}(1995){Caligari}, {Moreno-Insertis}, \&
  {Sch\"ussler}}]{CMS95}
{Caligari}, P., {Moreno-Insertis}, F., \& {Sch\"ussler}, M. 1995, \apj, 441,
  886

\bibitem[{{Cheung} {et~al.}(2010){Cheung}, {Rempel}, {Title}, \&
  {Sch{\"u}ssler}}]{Cheung_etal10}
{Cheung}, M.~C.~M., {Rempel}, M., {Title}, A.~M., \& {Sch{\"u}ssler}, M. 2010,
  \apj, 720, 233

\bibitem[{{Choudhuri}(2003)}]{Cho03}
{Choudhuri}, A.~R. 2003, \solphys, 215, 31

\bibitem[{{Guerrero} \& {K{\"a}pyl{\"a}}(2011)}]{GK11}
{Guerrero}, G., \& {K{\"a}pyl{\"a}}, P.~J. 2011, \aap, 533, A40

\bibitem[{{Haugen} \& {Brandenburg}(2004)}]{Hau04}
{Haugen}, N.~E.~L., \& {Brandenburg}, A. 2004, \pre, 70, 036408

\bibitem[{{Ilonidis} {et~al.}(2011){Ilonidis}, {Zhao}, \&
  {Kosovichev}}]{Ilonidis:2011}
{Ilonidis}, S., {Zhao}, J., \& {Kosovichev}, A. 2011, Sci, 333, 993

\bibitem[{{Isobe} {et~al.}(2005){Isobe}, {Miyagoshi}, {Shibata}, \&
  {Yokoyama}}]{Shibata05}
{Isobe}, H., {Miyagoshi}, T., {Shibata}, K., \& {Yokoyama}, T. 2005, \nat, 434,
  478

\bibitem[{{Jabbari} {et~al.}(2013){Jabbari}, {Brandenburg}, {Kleeorin},
  {Mitra}, \& {Rogachevskii}}]{JBKMR13}
{Jabbari}, S., {Brandenburg}, A., {Kleeorin}, N., {Mitra}, D., \&
  {Rogachevskii}, I. 2013, \aap, 556, A106

\bibitem[{{K{\"a}pyl{\"a}} {et~al.}(2013){K{\"a}pyl{\"a}}, {Mantere}, {Cole},
  {Warnecke}, \& {Brandenburg}}]{KMCWB13}
{K{\"a}pyl{\"a}}, P.~J., {Mantere}, M.~J., {Cole}, E., {Warnecke}, J., \&
  {Brandenburg}, A. 2013, ApJ, in press (arXiv:1301.2595)

\bibitem[{{Kemel} {et~al.}(2012){Kemel}, {Brandenburg}, {Kleeorin}, {Mitra}, \&
  {Rogachevskii}}]{KBKMR12}
{Kemel}, K., {Brandenburg}, A., {Kleeorin}, N., {Mitra}, D., \& {Rogachevskii},
  I. 2012, \solphys, 280, 321

\bibitem[{{Kitchatinov} \& {Mazur}(2000)}]{KM00}
{Kitchatinov}, L.~L., \& {Mazur}, M.~V. 2000, \solphys, 191, 325

\bibitem[{{Kleeorin} {et~al.}(1996){Kleeorin}, {Mond}, \&
  {Rogachevskii}}]{KMR96}
{Kleeorin}, N., {Mond}, M., \& {Rogachevskii}, I. 1996, \aap, 307, 293

\bibitem[{{Kleeorin} \& {Rogachevskii}(1994)}]{KR1994}
{Kleeorin}, N., \& {Rogachevskii}, I. 1994, Phys. Rev. E, 50, 2716

\bibitem[{{Kleeorin} {et~al.}(1990){Kleeorin}, {Rogachevskii}, \&
  {Ruzmaikin}}]{KRR1990}
{Kleeorin}, N., {Rogachevskii}, I., \& {Ruzmaikin}, A. 1990, JETP, 70, 878

\bibitem[{{Kleeorin} {et~al.}(1989){Kleeorin}, {Rogachevskii}, \&
  {Ruzmaikin}}]{Kleeorin89}
{Kleeorin}, N.~I., {Rogachevskii}, I.~V., \& {Ruzmaikin}, A.~A. 1989, PAZh, 15,
  639

\bibitem[{{Losada} {et~al.}(2012){Losada}, {Brandenburg}, {Kleeorin}, {Mitra},
  \& {Rogachevskii}}]{LBKMR2012}
{Losada}, I.~R., {Brandenburg}, A., {Kleeorin}, N., {Mitra}, D., \&
  {Rogachevskii}, I. 2012, \aap, 548, A49

\bibitem[{{Losada} {et~al.}(2013{\natexlab{a}}){Losada}, {Brandenburg},
  {Kleeorin}, \& {Rogachevskii}}]{LBKR2013a}
{Losada}, I.~R., {Brandenburg}, A., {Kleeorin}, N., \& {Rogachevskii}, I.
  2013{\natexlab{a}}, \aap, 556, A83

\bibitem[{{Losada} {et~al.}(2013{\natexlab{b}}){Losada}, {Brandenburg},
  {Kleeorin}, \& {Rogachevskii}}]{LBKR2013b}
{Losada}, I.~R., {Brandenburg}, A., {Kleeorin}, N., \& {Rogachevskii}, I.
  2013{\natexlab{b}}, A\&A, submitted (arXiv:1307.4945)

\bibitem[{{Nandy} \& {Choudhuri}(2001)}]{NC01}
{Nandy}, D., \& {Choudhuri}, A.~R. 2001, \apj, 551, 576

\bibitem[{{Nelson} {et~al.}(2013){Nelson}, {Brown}, {Brun}, {Miesch}, \&
  {Toomre}}]{NBBMT13}
{Nelson}, N.~J., {Brown}, B.~P., {Brun}, A.~S., {Miesch}, M.~S., \& {Toomre},
  J. 2013, \apj, 762, 73

\bibitem[{{Parker}(1955)}]{P55b}
{Parker}, E.~N. 1955, \apj, 121, 491

\bibitem[{{Parker}(1975)}]{Par75}
{Parker}, E.~N. 1975, \apj, 198, 205

\bibitem[{{Rempel} {et~al.}(2009){Rempel}, {Sch{\"u}ssler}, \&
  {Kn{\"o}lker}}]{Rempel09}
{Rempel}, M., {Sch{\"u}ssler}, M., \& {Kn{\"o}lker}, M. 2009, \apj, 691, 640

\bibitem[{{Rogachevskii} \& {Kleeorin}(2007)}]{RK2007}
{Rogachevskii}, I., \& {Kleeorin}, N. 2007, Phys. Rev. E, 76, 056307

\bibitem[{{Scharmer} {et~al.}(2011){Scharmer}, {Henriques}, {Kiselman}, \& {de
  la Cruz Rodr{\'{\i}}guez}}]{SHKR11}
{Scharmer}, G.~B., {Henriques}, V.~M.~J., {Kiselman}, D., \& {de la Cruz
  Rodr{\'{\i}}guez}, J. 2011, Science, 333, 316

\bibitem[{{Spruit}(2012)}]{Spruit12}
{Spruit}, H. 2012, PThPS, 195, 185

\bibitem[{{Stein} \& {Nordlund}(2012)}]{SN12}
{Stein}, R.~F., \& {Nordlund}, {\AA}. 2012, \apjl, 753, L13

\bibitem[{{Tao} {et~al.}(1998){Tao}, {Weiss}, {Brownjohn}, \&
  {Proctor}}]{Tao_etal98}
{Tao}, L., {Weiss}, N.~O., {Brownjohn}, D.~P., \& {Proctor}, M.~R.~E. 1998,
  \apjl, 496, L39

\bibitem[{{Warnecke} \& {Brandenburg}(2010)}]{WB10}
{Warnecke}, J., \& {Brandenburg}, A. 2010, \aap, 523, A19

\bibitem[{{Warnecke} {et~al.}(2011){Warnecke}, {Brandenburg}, \&
  {Mitra}}]{WBM11}
{Warnecke}, J., {Brandenburg}, A., \& {Mitra}, D. 2011, \aap, 534, A11

\bibitem[{{Warnecke} {et~al.}(2012{\natexlab{a}}){Warnecke}, {Brandenburg}, \&
  {Mitra}}]{WBM12}
{Warnecke}, J., {Brandenburg}, A., \& {Mitra}, D. 2012{\natexlab{a}}, JSWSC, 2,
  A11

\bibitem[{{Warnecke} {et~al.}(2012{\natexlab{b}}){Warnecke}, {K{\"a}pyl{\"a}},
  {Mantere}, \& {Brandenburg}}]{WKMB12}
{Warnecke}, J., {K{\"a}pyl{\"a}}, P.~J., {Mantere}, M.~J., \& {Brandenburg}, A.
  2012{\natexlab{b}}, \solphys, 280, 299

\bibitem[{{Warnecke} {et~al.}(2013){Warnecke}, {K{\"a}pyl{\"a}}, {Mantere}, \&
  {Brandenburg}}]{WKMB13b}
{Warnecke}, J., {K{\"a}pyl{\"a}}, P.~J., {Mantere}, M.~J., \& {Brandenburg}, A.
  2013, ApJ, in press (arXiv:1301.2248)

\end{thebibliography}
\bibliographystyle{apj}

\end{document}